\documentclass[prb,twocolumn,noshowpacs,amsmath,twoside]{revtex4}
\usepackage{bm}
\usepackage{epsfig}
\begin{document}

\title{Quantum suppression of shot noise in field emitters}

\author{O. M. Bulashenko}
 \email{oleg@ffn.ub.es}
\author{J. M. Rub\'{\i}}
\affiliation{Departament de F\'{\i}sica Fonamental,
Universitat de Barcelona, Diagonal 647, E-08028 Barcelona, Spain}

\date{30 September 2002}

\begin{abstract}
We have analyzed the shot noise of electron emission under strong applied 
electric fields within the Landauer-B\"uttiker scheme.
In contrast to the previous studies of vacuum-tube emitters, we show that
in new generation electron emitters, scaled down to the nanometer dimensions, 
shot noise much smaller than the Schottky noise is observable.
Carbon nanotube field emitters are among possible candidates to observe 
the effect of shot-noise suppression caused by quantum partitioning.
\end{abstract}


\maketitle

\section{Introduction}

Almost a century ago, Schottky pointed out that if electrons are emitted 
as discrete particles independently of each other, current fluctuations 
are to be expected with the noise power: $S_I=2qI$, with $q$ the elementary 
charge and $I$ the mean current. \cite{schottky18}
This phenomenon, called the ``shot effect'' by Schottky, 
was later observed in vacuum tubes \cite{hartmann21,hull25}
in nice agreement with his prediction.

During the last two decades, the shot effect (now called shot noise) 
has been discovered and intensively studied 
in mesoscopic phase-coherent conductors. \cite{blanter00}
In a quantum point contact (QPC), for instance, the current-noise power 
was found to be $S_I=2qI\,(1-{\cal T})$, where ${\cal T}$ is the transmission 
probability (for one-channel transmission). 
In this formula, the noise is suppressed by the factor $1-{\cal T}$ 
relative to the Schottky result, thereby predicting zero noise for perfect 
transmission (see experimental evidence 
\cite{reznikov95,kumar96,vandenbrom99}).
In both cases, in QPCs and vacuum tubes, the granularity of charge is 
manifested in the shot noise, although the source of randomness is different:
\cite{oberholzer02}
In QPC, the randomness appears in the transmission process due to 
the quantum partitioning between the incoming and outgoing states 
(the incoming carriers are noiseless).
In contrast, in vacuum tubes, the randomness is an inherent property
of the emitter caused by thermal fluctuations.

An interesting question then arises: Is it possible to observe 
the quantum partition noise in electron emission, in the same way as in QPCs,
with the noise power suppressed below the $2qI$ value? 
The related question---whether the shot noise in Schottky's 
vacuum tube is classical---has been addressed recently by 
Sch\"onenberger, Oberholzer, Sukhorukov, and Grabert.
\cite{schonenberger01,oberholzer01}
The authors showed that for the vacuum tube parameters 
typical for the earlier stages of development of vacuum electronics, 
\cite{hull25} the quantum partitioning in electron emission is absent,
and consequently the shot noise observed in Schottky's vacuum tubes is
classical. 

In this paper we show that in new generation electron emitters,
scaled down to the nanometer dimensions, shot noise much smaller than 
the Schottky noise, due to a quantum partitioning effect, is observable. 
Moreover, two different sources of randomness---thermal agitations and
quantum partitioning---may act together governing the electron emission noise.
\cite{remark0}
A rapidly growing field of nanoscale electronics suggests to us various 
examples of electron emitters in which this phenomenon may be tested: 
the nanotube field emitters, \cite{saito99,fransen99,cumings02} 
the composite emitters coated by wide-band-gap, low-work-function, 
and/or negative-electron-affinity materials, \cite{modukuru00,binh00,sugino02} 
diamondlike emitters, \cite{geis98,auciello01} among others. 

\section{Shot noise in electron emission}

We start by considering the electron emission as a quantum scattering problem 
within the Landauer-B\"uttiker framework. 
The transverse and longitudinal motion of electrons are assumed to be 
separable, so that one can specify the quantum channels associated with
transverse modes, and define the scattering states.
The equation for the mean current in a phase-coherent conductor attached to 
two electron reservoirs with different chemical potentials reads
\cite{buttiker92}
\begin{equation}  \label{mcur2}
I = \frac{q}{\pi\hbar} \int d\varepsilon\
(f_L-f_R)\, {\rm Tr}({\bf t}^{\dag}{\bf t}),
\end{equation}
where $f_{L,R}(\varepsilon)$ are the energy distribution functions 
at the left ($L$) and right ($R$) reservoirs, 
${\bf t}$ is the matrix of the transmission amplitudes,
\cite{buttiker92,blanter00} and the trace 
is taken over all the transmission channels at energy $\varepsilon$.
For definiteness, the left reservoir is considered as an emitter, 
and the right reservoir, to which an external positive bias is applied, 
as a collector. We assume that the ``quantum conductor'' between 
the two reservoirs could also be a vacuum gap.
At the surface of the emitter---between the emitter-vacuum or 
emitter-semiconductor interface---a potential barrier exists, which limits
the current and scatters the emitted electrons 
(only elastic scattering is assumed). 
Thus the transmission matrix ${\bf t}$ is referred to the scattering 
on the potential barrier. 

The zero-frequency current-noise power for a two-terminal quantum conductor 
is given by \cite{buttiker92,kogan}
\begin{multline} \label{Stwo2}
S_I = 2G_0 \int d\varepsilon\ \{ [f_L (1-f_L) + f_R (1-f_R)] \,
{\rm Tr}({\bf t}^{\dag}{\bf t}{\bf t}^{\dag}{\bf t}) \\
+ [f_L (1 - f_R) + f_R (1 - f_L)] \, [{\rm Tr}({\bf t}^{\dag}{\bf t}) - 
{\rm Tr}({\bf t}^{\dag}{\bf t}{\bf t}^{\dag}{\bf t})] \},
\end{multline}
with $G_0=q^2/\pi\hbar$ the unit of conductance.
For sufficiently high biases, all the states in the collector at energies
corresponding to the occupied states at the emitter 
(that contribute to the emission) are empty. Hence one can take $f_R=0$. 
In this case, the steady-state emission current 
in the basis of eigen-channels becomes
\begin{equation}  \label{mcur1}
I = \frac{q}{\pi\hbar} \sum_n \int d\varepsilon\ f \, {\cal T}_n,
\end{equation}
where ${\cal T}_n$ are the transmission probabilities associated with 
$n$ quantum channels at energy $\varepsilon$. 
Hereafter, we drop the subindex $L$ at the occupation numbers $f$, since only
the emitter contact contributes to the current and noise.
The noise power (\ref{Stwo2}) for the unidirectional injection becomes
\begin{align} \label{Sone}
S_I &= 2G_0 \sum_n \int d\varepsilon\ [
f (1-f) \, {\cal T}_n^2  +  f \, {\cal T}_n (1-{\cal T}_n)  ] \nonumber \\
&\equiv S_I^{\rm em} + S_I^{\rm part}.
\end{align}
This formula describes the spectral density of current fluctuations 
of an electron emitter. 
It unifies two sources of randomness: 
(i) the probabilistic occupation of states in the emitter
(through the function $f$) and 
(ii) the probabilistic reflection and transmission at the interface barrier
(through the probabilities $T_n$). 
The first source of randomness is intimately related to intrinsic thermal 
agitations of the emitter
and can be associated with the first term in Eq.~(\ref{Sone}).
Since $f(1-f)=-k_BT (\partial f/\partial \varepsilon)$, this term
is related to a thermal broadening of the occupation numbers 
at the Fermi level.
Note that it vanishes at zero temperature, but dominates in the absence of 
partitioning when all transmission coefficients  $T_n$ are either $0$ or $1$,
and hence can be interpreted as the {\em emission shot noise}.
The second source of randomness associated with the last term 
in Eq.~(\ref{Sone}) is caused by quantum partitioning 
and the fact that charge is carried by discrete portions (shot effect).
It only contributes for transmission probabilities ${\cal T}_n\neq 0,1$,
it does not vanish at zero temperature, and can be called the 
{\em partition shot noise}.
It is clear that both noise sources act together and cannot be separated,
in general. 
\cite{schonenberger01,oberholzer01}
For future analysis, it is convenient, however, to introduce the notations 
for the emission noise $S_I^{\rm em}$ and the partition noise $S_I^{\rm part}$
according to the above discussion.

Equation (\ref{Sone}) can be used to calculate the noise power
of the emitter with an arbitrary number of quantum channels.
The problem can be simplified by assuming that the interface of the emitter 
is plane and its transversal area is large compared with wavelength
(a large number of channels). Then, the summation over the transverse 
channels can be replaced by integration over
the transversal energy $E_{\perp}$, which can be performed giving
\begin{multline} \label{SI_FD}
S_I = 2G_S \frac{k_BT}{E_F} \left\{ 
\int_0^{\infty}  \frac{{\cal T}^2(E)}{1+e^{(E-E_F)/k_BT}} dE \right. \\ \left. 
+ \int_0^{\infty} {\cal T}(E) [1-{\cal T}(E)]\,
\ln[1+e^{(E_F-E)/k_BT}]\, dE \right\},
\end{multline}
where $G_S=(k_F^2 A/4\pi)\,G_0$ is the Sharvin conductance, 
$E_F=\hbar^2 k_F^2/(2m)$ is the Fermi energy, $A$ is the cross sectional area,
and ${\cal T}(E)$ is the transmission probability at the longitudinal energy
$E=\varepsilon-E_{\perp}$.
Now, we can verify Eqs.~(\ref{Sone}) and (\ref{SI_FD}) for two practical 
cases: thermionic emission and field emission.

\subsection{Thermionic emission}

When the potential barrier is wide on the scale of the wavelength,
one can neglect tunneling. 
In this case, an appreciable emission current
can be achieved, for instance, by heating the emitter, so that thermally 
excited electrons escape above the barrier.
The transmission probability takes the values 1 for $E>\Phi_b$
and 0 for $E<\Phi_b$, where $\Phi_b$ is the barrier height 
(quantum reflection for overbarrier electrons is negligible for a sufficiently
smooth potential). 
Thus the partition term vanishes and the noise contains only the emission 
(thermionic) contribution:
\begin{equation} \label{Sone_em}
S_I \approx S_I^{\rm em} = 2G_0 \sum_{n^*} \int d\varepsilon\ f (1-f),
\end{equation}
where the summation is taken for open channels only.
For wide multichannel emitters with equilibrium Fermi-Dirac electrons 
[see Eq.~(\ref{SI_FD})], Eq.~(\ref{Sone_em}) is reduced to 
\begin{equation} \label{SI_em}
S_I = 2G_S \frac{(k_BT)^2}{E_F}\,
\ln[1+e^{(E_F-\Phi_b)/k_BT}].
\end{equation}
This formula gives the $2qI$ Schottky value, whenever $E_F< \Phi_b-3k_BT$,
which is a condition for a nondegenerate Maxwell-Boltzmann injection
(Richardson-Laue-Dushman regime of thermionic emission \cite{modinos}).
For a degenerate injection, $E_F\agt\Phi_b$, the noise is suppressed 
below the Schottky value by the factor \cite{prb01}
${\cal F}_0(\zeta)/{\cal F}_1(\zeta)$, where
$\zeta=(E_F-\Phi_b)/k_BT$, and ${\cal F}_k$ is the Fermi-Dirac integral
of index $k$.
This suppression is caused by Fermi correlations imposed by the Pauli exclusion
principle (see Ref.~\onlinecite{prb01} for the details).
Note that for metallic cathodes used in vacuum tubes, \cite{hull25}
the work function is about 4 eV, which is much larger with respect
to $k_BT$, so that only nondegenerate injection with the full shot noise 
is possible, \cite{remark2} as was observed in the experiment. \cite{hull25}

\subsection{Field emission}

The potential barrier at the emitter can be narrowed by applying
a strong electric field, so that electrons can be pulled out from the cold 
emitter via quantum tunneling. \cite{fowler28}
In this case, the partition noise is expected to be the dominating source 
of noise:
\begin{equation} \label{Sone_part}
S_I \approx S_I^{\rm part} 
= 2G_0 \sum_n \int d\varepsilon\ f \, {\cal T}_n (1-{\cal T}_n).
\end{equation}
By applying this formula again to Fermi-Dirac electrons in a wide emitter
under the condition $k_BT\ll E_F$, 
we obtain the noise which is independent of temperature,
\begin{align} \label{SI_FD0}
S_I = 2G_S \int_0^{E_F} {\cal T}(E) [1-{\cal T}(E)]\,
\left(1-\frac{E}{E_F}\right)\, dE.
\end{align}
In the Fowler-Nordheim regime of field emission, \cite{fowler28} 
when the Fermi energy of the emitter is much below the barrier top, 
the transmission probability for electrons at the Fermi level 
(which mostly contribute to the emission) is small, \cite{modinos} 
${\cal T}\ll 1$. It can be verified that in this regime, 
Eq.~(\ref{SI_FD0}) gives the Schottky $2qI$ law.

\subsection{Poissonian versus non-Poissonian emission}

Summarizing these two examples, one can conclude that the Schottky noise,
which is the noise produced by independently injected electrons
(Poissonian process), may occur under two physically different conditions: 
\cite{blanter00}
(i) the low occupation numbers, $f\ll 1$, when electrons are initially
Poissonian and remain Poissonian after passing the barrier
with whatever probability; (ii) the low transmission probability, 
${\cal T}\ll 1$, when the incoming electrons may be initially noiseless,
but after tunneling through the barrier, the outgoing flow becomes diluted and 
obeys a Poissonian statistics. 
Although in both cases, the noise power is given by the Schottky law 
$S_I=2qI$, in the former case its value is sensitive to the temperature,
while in the latter case it is not. This fact may be used in the experiment
to distinguish these two mechanisms.

Now it is clear under which conditions one should expect a deviation
from the $2qI$ law. It is the case when both the occupation numbers $f$
and the transmission probabilities ${\cal T}$ are not small with respect to 1.
This is precisely the situation that may occur in novel field emitters. 
The requirements of strong electric currents under low voltages
led research interests towards low-work-function materials 
(low potential barriers) and sharp emitter tips (narrow potential barriers). 
For instance, extremely high electric fields $\sim 10^8\,
{\rm V/cm}$
are achieved in nanotube emitters due to a geometric field enhancement
in high-aspect-ratio tips. \cite{collins96}
For combinations of work function, field, and temperature parameters in
many of these emitters, an appreciable part of electron emission originates 
from energy levels in the vicinity (below and above) of the potential barrier. 
To estimate the noise in this case, one should use general formulas given by 
Eq.~(\ref{Sone}) or (\ref{SI_FD}), in which the transmission coefficients 
must be known in a wide energy range,---below and above the barrier.

\section{Shot noise for emission through a triangular barrier}

In this paper, we consider the model that consists of a simplified
triangular representation of the barrier with only two parameters: the height 
of the barrier $\Phi_b$ and the slope determined by the electric field $F$:
\begin{equation} \label{Vx}
V(x) = \begin{cases} 0, & x<0 \\ \Phi_b-qFx, & x\geq 0. \end{cases}
\end{equation}
The potential barrier height $\Phi_b$ is equal to the electron affinity for 
semiconductors, and for metals it is a sum of the work function and
the chemical potential. \cite{modinos} 
We neglect the rounding off the barrier due to the image interaction.
Although this may cause some quantitative error, 
such a consideration enables the exact solution of the Schr\"odinger equation
in terms of Airy functions and thus an exact evaluation of the electron 
transmission probability in the whole energy range including the barrier top,
which is not allowed in the WKB scheme. \cite{remark3}
The transmission probability ${\cal T}(E)$ for an arbitrary incident energy $E$
(below and above the barrier top) can be represented by a unique analytical
formula (see the Appendix).
\begin{figure}[t]
\epsfxsize=8.0cm
\epsfbox{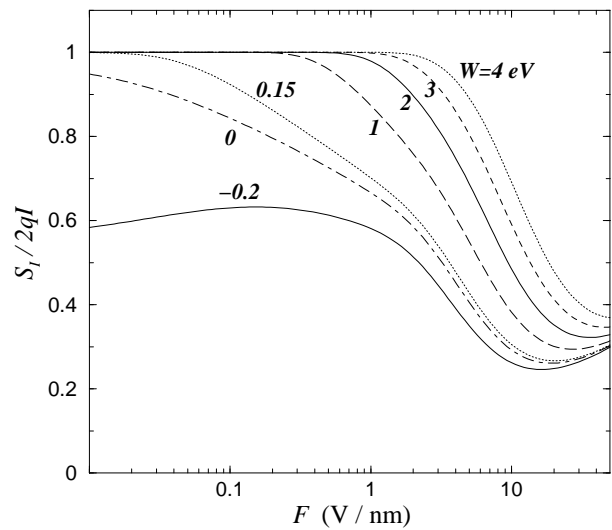}
\caption{Current-noise power $S_I$ normalized to the Schottky value $2qI$
as a function of the electric field $F$ for electron emission through 
a triangular potential barrier. The curves are plotted 
for various potential barrier heights (i.e., work functions $W=\Phi_b-E_F$,
with $E_F$=4 eV).}
\label{fsi-F} \end{figure}

To illustrate the results for the current emission noise, consider 
the emitter at $T$=300 K with $E_F=4\,{\rm eV}$ in which, for simplicity, 
the electron effective mass for the emitter and the barrier is the same. 
The noise spectral densities $S_I$ calculated from Eq.~(\ref{SI_FD}) 
are plotted in Fig.~\ref{fsi-F} as functions of the electric field $F$
for different barrier heights (work functions) $W=\Phi_b-E_F$.
The zero-field limit $F\to 0$ corresponds to the emission over
a rectangular-step barrier with no tunneling, for which 
${\cal T}(E<\Phi_b)\to 0$.
In this limit, the noise power is determined by the relative position of
the Fermi level with respect to the barrier top. 
For $W\agt 0.15$~eV, the injection is nondegenerate and, consequently, 
$S_I^{\rm em}\propto\int_{\Phi_b}^{\infty}{\cal T}^2e^{-E/k_BT}dE$ and
$S_I^{\rm part}\propto\int_{\Phi_b}^{\infty}{\cal T}(1-{\cal T})e^{-E/k_BT}dE$.
Their sum gives the full Schottky noise:
$S_I\propto\int_{\Phi_b}^{\infty}{\cal T}e^{-E/k_BT}dE=2qI$. 
Since the transmission probabilities are not exactly equal to 1 for all the 
energies $E>\Phi_b$ (there is a finite reflection due to a sharp potential 
change), both terms contribute to the noise. 
It is also seen from Fig.~\ref{fsi-F} that for $W\alt0.15$~eV, the noise 
at low fields is expected to be smaller than the Schottky value. 
This is the case of a degenerate injection, for which the quadratic terms 
$\sim {\cal T}^2$ in the two contributions do not cancel, in contrast
to the previous case of a nondegenerate injection.
The noise suppression effect here originates from the Fermi statistical
correlations under the condition of current partitioning.

Although the case $W<0.15$~eV for the field emitters is likely to occur, 
\cite{binh00,gohda00}
the most typical case in practice is the opposite condition $W>0.15$~eV.
In this case, at low fields, the current noise is the full Schottky noise. 
Our prediction is that when the electric field increases, there exists 
a threshold value, at which the noise starts to drop down 
the Schottky value (see Fig.~\ref{fsi-F}). 
It can be explained by the fact that as the electric field 
increases, the barrier becomes thinner, the transmission probability increases,
and the noise starts to drop down due to the $1-{\cal T}$ factor 
[see Eq.~(\ref{SI_FD0})].
It is worth noting that this drop is a quantum phenomenon. 
The quantum uncertainty of whether the electron has been transmitted through 
or reflected by the barrier is a source of randomness which produces 
the partition noise $\propto{\cal T}(1-{\cal T})$.
For pure classical transmission, when the probabilities of transmission at 
different energies are either 0 or 1, the quantum partition noise does not 
appear and the noise of the emitted electrons is governed by the first term 
on the right-hand side of Eq.~(\ref{SI_FD}) which gives the $2qI$ law with 
no drop. 
We would like to highlight that the suppression effect caused by quantum 
partitioning is independent of the degeneracy of electrons. Even for a fully 
degenerate case at zero temperature, for which electrons initially are 
noiseless, after passing the barrier they acquire a partition noise 
with a suppression level sensitive to the transmission coefficient. 

The threshold value for the electric field, for which the shot-noise 
suppression becomes clearly visible, may be roughly estimated from 
Eq.~(\ref{TE}) by taking the value for the tunneling probability at the Fermi 
energy to be ${\cal T}(E_F)\approx 0.1$. It is seen from Fig.~\ref{fsi-F} 
that for the work function 4 eV, this threshold field is about 
$3\, {\rm V/nm}$, for the work function 3 eV, it is $\sim 2\,{\rm V/nm}$, 
and for 2 eV, it is $\sim 1\,{\rm V/nm}$,
values that do not seem unrealistic.  \cite{fransen99,cumings02}
The threshold field may also be decreased by choosing the emitter with
a high effective mass of electrons (e.g., of heavy-fermion materials), since 
the transmission probability (\ref{TE}) is sensitive to the ratio 
of the effective masses.
The barrier lowering due to a self-consistent potential redistribution
may also decrease the threshold field value.
Note that for negative affinity materials, the quantum suppression of shot 
noise should be observed for arbitrarily small fields.

\section{Conclusions}
In conclusion, we have calculated the shot noise power of electron emission 
under the action of strong applied electric fields.
Within the Landauer-B\"uttiker scheme, we have shown that the emission noise 
is governed by two different stochastic processes acting together---thermal 
agitations and quantum partitioning. 
The analytical formula for the noise power, which unifies these two sources 
of randomness, has been analyzed.
This formula, in the limit of a wide potential barrier (no tunneling), 
describes the shot noise of thermionic emission, 
which may be either Poissonian for a nondegenerate injection
(Richardson-Laue-Dushman regime), or non-Poissonian for a degenerate injection.
Under field-emission conditions, the noise recovers the full Schottky noise
in the Fowler-Nordheim regime. 

Our results indicate that in order to observe the shot-noise suppression
in field emitters below the Schottky level, 
there are at least two possibilities:
(i) by lowering the work function $W$; 
then, the noise starts to be sub-Poissonian below some value $W$ 
at arbitrarily small fields;
(ii) for high-work-function materials, by increasing the electric field $F$
at the emitter tip, e.g., by employing the nanotube emitters;
then, the noise starts to be sub-Poissonian above some threshold value $F$.
Note that precisely in the regime when the shot noise deviates from the 
Schottky law, the Fowler-Nordheim plots [$\ln(I/F^2)$ vs $1/F$]
no longer follow the straight lines. \cite{modinos,fransen99} 
The measurements of the noise suppression value (with respect to the Poissonian
value) may provide additional data on the work function and the electric field 
at the emitter tip---important information not available from the 
current-voltage characteristics alone (especially for new unknown materials).
\cite{fransen99}
Since the noise is sensitive to the injection energy profile, \cite{pe02}
the noise measurements may serve as a substitution for direct field-emission 
energy profile measurements.

It is clear why the quantum partitioning has not been observed in noise 
measurements on metal-cathode vacuum tubes a long time ago, at the earlier 
stages of the development of vacuum electronics. \cite{hull25}
In those experiments, the electric-field values at the emitter were no more 
than 0.003 V/nm, which is too low to see the effect on 4 eV work function 
materials. We are not aware of shot-noise measurements under field emission at 
stronger fields.
The typical values obtained from old literature \cite{modinos}
indicate fields of about 2---5 V/nm, and in novel nanotube field emitters
they are even greater. \cite{fransen99,cumings02}
We believe that in such conditions, the quantum suppression of shot noise
is observable.
Besides the nanotube emitters, the suitable candidates to observe
the shot-noise suppression could be the composite emitters coated by 
wide-band-gap, low-work-function, and/or negative-electron-affinity materials, 
\cite{modukuru00,binh00,sugino02} 
or diamond-like emitters. \cite{geis98,auciello01}

\begin{acknowledgements}
We thank Eugene Sukhorukov for valuable discussions.
This work has been partially supported by the Ministerio de Ciencia y 
Tecnolog\'{\i}a of Spain through the ``Ram\'on y Cajal'' program,
and by the DGICYT and FEDER under Grant No.\ BFM2002-01267.
\end{acknowledgements}

\appendix
\section{Transmission probability}

The transmission coefficient can be found as a solution of the scattering 
problem for a one-dimensional potential barrier,
since it depends only on the energy of the longitudinal motion $E$.
The time-independent Schr\"odinger equation is given by
\begin{equation} \label{Schrod}
-\frac{\hbar^2}{2m^*} \frac{d^2\psi}{dx^2} + V(x)\, \psi = E \, \psi,
\end{equation}
in which the potential $V$ is defined by Eq.~(\ref{Vx}).
We assume that, in general, the electron effective mass $m^*$ 
in Eq.~(\ref{Schrod}) may differ for the emitter and the barrier:
$m^*=m_e$ for $x<0$ and $m^*=m_b$ for $x>0$.

The solutions of the Schr\"odinger equation for the constant potential are
the plain waves, while for the linear potential the solutions are given
by the Airy functions:
\begin{equation}\begin{split}
\label{psi}
\psi(x<0) &= a\ e^{ikx} + b\ e^{-ikx}, \\
\psi(x>0) &= t\, \left\{
{\rm Bi}\, [k_A(w-x)] + i\,{\rm Ai}\, [k_A(w-x)] \right\},
\end{split}\end{equation}
where $k=\sqrt{2m_e E}/\hbar$ is the momentum of the incident electron,
$k_A=(2m_b qF/\hbar^2)^{1/3}$ is the characteristic momentum in the arguments 
of the Airy functions dependent on the field, 
and $w=(\Phi_b-E)/(qF)$ is the coordinate where the Airy functions change 
from monotonic to oscillatory behavior (for positive values, 
$w$ is just the barrier width at energy $E$).
Since we calculate the transmission, the solution $\psi(x>0)$ corresponds 
to the outgoing wave at $x\to\infty$.
\cite{remark4}

{}From the continuity of the wave function and the current conservation
at the interface, we obtain the transmission amplitude
\begin{equation} \label{tt}
t = \frac{4a}{{\rm Bi}(z)- \sigma\, {\rm Ai}'(z)
+ i\,[{\rm Ai}(z) + \sigma\, {\rm Bi}'(z)]},
\end{equation}
where $\sigma=(k_A/k)(m_e/m_b)$ and 
the prime on the Airy functions indicates a derivative with respect to 
the argument $z=k_A w$.
The incident and transmitted current densities corresponding to the
wave functions (\ref{psi}) are found as $j_{\rm inc}=(\hbar k/m_e)|a|^2$ and 
$j_{\rm trans}=(\hbar k_A/\pi m_b) |t|^2$.
Therefore, the transmission coefficient for a given incident momentum $k$ is
${\cal T} = j_{\rm trans}/j_{\rm inc}=(k_A m_e/\pi k m_b)|t|^2/|a|^2$,
which finally gives
\begin{widetext}
\begin{equation} \label{TE}
{\cal T}(E) = \frac{4}{2 + (\pi/\sigma)\,[{\rm Ai}^2(z) + {\rm Bi}^2(z)]
+ \pi\sigma\, [{\rm Ai}'^2(z) + {\rm Bi}'^2(z)]},
\end{equation}
\end{widetext}
in which the energy dependence appears through both the momentum $k$ 
in the parameter $\sigma$ and 
the argument of the Airy functions $z=2m_b(\Phi_b-E)/(\hbar^2 k_A^2)$.


\begin{thebibliography}{}
\bibitem{remark0}
We consider the noise at frequencies beyond the $1/f$ spectrum.
\bibitem{remark3}
It should be noted that the WKB evaluation of the transmission coefficient,
traditionally used to calculate the emission currents, 
is not suitable for the calculation of the partition noise.
The WKB gives an appreciable error in the vicinity of the classical
turning points, in particular, for electron energies near the barrier top
(see Ref.~\onlinecite{modinos}).
\bibitem{remark4}
Strictly speaking, in order to avoid contradiction with the definition 
of the scattering states, the boundary conditions for the Airy functions 
must also be specified at the collector.
It is assumed, however, that the applied bias is sufficiently high, so that
the potential details at the collector do not affect
the electron tunneling at the emitter.
\end{thebibliography}
\end{document}